\documentclass[12pt]{article}
\usepackage{amsmath, euscript, amssymb, amsfonts}

\usepackage{amsthm}

\newtheorem{theorem}{Theorem}
\newtheorem{lemma}{Lemma}

\theoremstyle{definition}
\newtheorem{definition}{Definition}

\theoremstyle{remark}
\newtheorem{remark}{Remark}

\newcommand{\oR}{{\mathbb R}}
\newcommand{\oV}{{\mathbb V}}
\newcommand{\oC}{{\mathbb C}}
\newcommand{\oZ}{{\mathbb Z}}

\newcommand{\oT}{{\mathbb T}}

\newcommand{\oJ}{{\mathbb J}}

\newcommand{\supp}{\mathop{\rm supp}\nolimits}

\newcommand{\I}{\mathop{\rm Im}\nolimits}

\newcommand{\gA}{{\mathfrak A}}
\newcommand{\gB}{{\mathfrak B}}
\newcommand{\gM}{{\mathfrak M}}
\newcommand{\ga}{{\mathfrak a}}
\newcommand{\gb}{{\mathfrak b}}
\newcommand{\gc}{{\mathfrak c}}
\newcommand{\gote}{{\mathfrak e}}
\newcommand{\gm}{{\mathfrak m}}

\newcommand{\be}{\begin{equation}}
\newcommand{\ee}{\end{equation}}
\newcommand{\bt}{\begin{theorem}}
\newcommand{\et}{\end{theorem}}

\begin{document}
\title{
 \normalsize\bf  PCT,  SPIN AND STATISTICS, AND ANALYTIC WAVE FRONT SET}
\date{}
\author{M.~A.~Soloviev\footnote{E-mail: soloviev@lpi.ru}}
\maketitle
\vspace{-5mm}

 \centerline{\sl P.~N.~Lebedev Physical Institute}
 \centerline{\sl Russian Academy of Sciences}
 \centerline{\sl  Leninsky Prospect 53, Moscow 119991, Russia}

\vskip 3em
\begin{abstract}
A new, more general derivation of the spin-statistics and PCT
theorems is presented. It uses the notion of the analytic wave
front set of (ultra)distributions and, in contrast to the usual
approach, covers nonlocal quantum fields. The fields are defined
as generalized functions with test functions of compact support in
momentum space. The vacuum expectation values are thereby admitted
to be arbitrarily singular in their space-time dependence. The
local commutativity condition is replaced by an asymptotic
commutativity condition, which develops generalizations of the
microcausality axiom previously proposed.
\end{abstract}

{\baselineskip=10pt

\tableofcontents}

\newpage
 \section[Introduction]{\large Introduction}

 In this paper, we examine how essential the locality of interaction and the
 microcausality axiom are for  deriving the two fundamental observed
 consequences of the general theory of quantum fields: the
 spin-statistics relation and the  PCT symmetry. We show that both of them hold if
 local commutativity is replaced with a
 condition which is closer related to macrocausality and could be called
 asymptotic commutativity. Intuitively, it implies that
 the commutators of observable fields  decrease for large spacelike
 separations of the arguments no slower than exponentially with order one and
with maximum type. The precise definition of this condition given
below is a refinement of that proposed in \cite{1}, where it was
compared with generalizations of local commutativity suggested by
 other authors. Abandoning microcausality, we  naturally eliminate the
 bounds on the high-energy (ultraviolet) behavior of the off-mass-shell
 amplitudes that follow from this axiom (see Sec.~9.1.D in \cite{2} and
 Sec.~VII.4 in \cite{3}).  In our setting, this behavior is arbitrary,  and
 so the space-time dependence of fields can be arbitrarily singular.
The usual derivation \cite{2}--\cite{4}
  of the  spin-statistics relation and  the  PCT symmetry, which
  is based on the analyticity properties of vacuum expectation values in
 coordinate space,  fails in this case, and
  an alternative construction
 of some envelopes of holomorphy in momentum space was
 suggested instead in  \cite{5}--\cite{7}. In \cite{8}, \cite{9},
 it was  observed that the
 problem can be solved  using  the notion of the analytic wave front
 set of distributions.
\par
 Here, we present all details  of the new derivation of
 the spin-statistics and   PCT theorems based on  this approach.
 Aiming for maximal generality, we choose  the Gelfand-Shilov spaces
  $S^0_\alpha$  as the functional
 domain of definition of fields. These spaces provide an enlarged framework
 as compared to the space $S^0$ used in \cite{5}--\cite{9},
  which is just the Fourier transform of Schwartz's space $\cal D$ of infinitely
  differentiable functions of compact support.
 In other words, the fields  under study are treated as
 ultradistributions rather than distributions in the momentum-space
 variables. Some recent results \cite{10}--\cite{13} in the theory
 of analytic functionals are essential for our
 derivation, in particular, that every functional of the class
 $S^{\prime\, 0}_\alpha$ has a unique minimal carrier cone.\footnote{From here on, the
 continuous dual space of a topological vector space is denoted by the same
 symbol with a prime.} Structure theorems
  concerning the properties of such a quasi-support allow
 handling highly singular generalized functions as easily as the
standard Schwartz distributions.

 The quantum theory of highly singular interactions is perhaps the most
 advanced branch of nonlocal field theory (see \cite{14}--\cite{17} for a review).
  A consistent
 theory of asymptotic states and particles was constructed within
 this  framework, including the derivation
 of high-energy bounds on the scattering amplitudes \cite{14}, \cite{18}.
  Using highly singular nonlocal
 form-factors proved efficient  for the phenomenological description
 of strong interactions \cite{16}. A possible interplay
 between this branch of field theory and  string theory is of particular interest
  (see \cite{1}, \cite{19}).
\par
 The work is organized as follows. Section~2  contains
  the above-mentioned structure theorems. In Sec.~3, a relationship between
  the analytic wave front set of ultradistributions and the carrier cones of
  their Fourier transforms is established. This relationship  shows the incompatibility of
  certain support properties in the coordinate and momentum spaces and is a key
  point in our approach. In  Sec. 4, the condition of asymptotic commutativity is
  formulated  as a restriction on the matrix elements of the field
  commutators and is then  rewritten in terms of the vacuum expectation
  values.  In Sec.~5,  the features of the invariant ultraviolet
  regularization of the Lorentz-covariant functionals of class
    $S^{\prime\, 0}_\alpha$ are investigated.  In Sec. 6, this regularization is
    used  to demonstrate the role of the Jost points in nonlocal
    field theory.  In  Sec.~7,  analogues of all main steps of the
    classical proof \cite{2}--\cite{4} of the spin-statistics theorem are derived for
    a    field theory subject to the  asymptotic commutativity condition,
    starting from the Dell'Antonio lemma  and finishing with the Araki
    theorem,
    which establishes the existence of the Klein transformation that reduces
    fields to the normal commutation relations. In Sec.~8, a condition
    of  weak  asymptotic commutativity is defined and its equivalence to
    the  PCT invariance is proved. We give some concluding
     remarks in Sec.~9.

 \section[Carrier cones of analytic functionals: Basic theorems]
 {\large Carrier cones of analytic functionals: Basic theorems}

In the standard axiomatic approach \cite{2}--\cite{4}, quantum
fields are assumed to be tempered operator-valued distributions
defined on the  test functions belonging to the Schwartz space $S$
because the singularities  occurring in the perturbation theory
are always of  finite order. However,  when we seek exact
solutions or consider general consequences of QFT, this assumption
proves  too restrictive (see \cite{20}). To generalize the
setting, we  use the Gelfand-Shilov spaces $S^\beta_\alpha$
\cite{21}. We recall that $S^\beta_\alpha(\oR^n)$ is the union
(more precisely, the inductive limit) of the family of spaces
$S^{\beta,b}_{\alpha,a}\quad (a,b > 0)$ consisting of those
infinitely differentiable functions on $\oR^n$ for which
 the norm
 \be \|f\|_{a,b}=\sup_{x\in \oR^n}\, \sup_{k,q\in
\oZ^n_+} \frac{|x^k\,\partial^q f(x)|}{a^{|k|}\,b^{|q|}\,k^{\alpha k}q^{\beta
q}}
 \label{(1)}
\ee
 is finite. Clearly, the index  $\alpha$ characterizes the growth rate
 of the functionals belonging to the dual space
$S^{\prime \,\beta}_\alpha$,  and $\beta$ characterizes their
singularity: in the QFT context, these indices control the
respective infrared and ultraviolet behavior of fields. In
accordance with the introduction, we set $\beta=0$ in what
follows. The space $S^0_\alpha$ is nontrivial only if $\alpha>1$,
which is assumed throughout this paper. The Fourier transformed
space $\mathcal F(S_\alpha^0)=S^\alpha_0$ consists of functions
with compact support. Therefore, the local
 properties of the generalized functions belonging to $S^{\prime\,\alpha}_0$,
which are called ultradistributions of the Roumieu class
 $\{k^{\alpha k}\}$, are the same as those of the Schwartz tempered
distributions, and their supports are defined in the standard way
through  a partition of unity. In contrast, the space $S^0_\alpha$
itself consists of entire analytic functions, i.e., in the
$x$-representation, the elements of its dual space are analytic
functionals for which the notion of a support is useless.
Nevertheless, the functionals of this class retain a kind of
angular localizability, which can be described as in \cite{10}. To
each open cone $U\subset \oR^n$, we assign a
 space $S^0_\alpha(U)$, which is defined similarly to  $S^0_\alpha$ with
 the only difference that in the analogue of  formula~\eqref{(1)}, the supremum
 must be taken over $x\in U$.  Every function  $f\in S^{0,b}_{\alpha,a}(U)$ allows
 an analytic continuation to the whole space $\oC^n$ and satisfies the estimate
 \be
 |f(x+iy)|\leq
 C\exp\{-|x/a'|^{1/\alpha}+b'd(x,U)+b'|y|\},
 \label{(2)}
 \ee
 where $d(\cdot,U)$ is the distance from the point to the cone $U$
 and where  $a'$ and $b'$ differ from $a$ and $b$  by a number factor that
  depends on the choice of
the norm\footnote{As a rule this norm is assumed to be Euclidean
in what follows.}
 in $\oR^n$.  This difference is unessential for the union  taken over
all the indices $a$ and $b$,  and  estimate~\eqref{(2)} can
therefore be used as a starting point for the definition of
$S^0_\alpha(U)$ as well. The spaces over cones possess the same
properties convenient from the standpoint of functional analysis
as those of the original $S^0_\alpha$. Namely, these spaces are
complete, barrelled, reflexive and Montel (see \cite{17}).  The
space $S^0_\alpha$ is evidently a linear subspace of every
$S^0_\alpha(U)$, and $S^{\prime\,0}_\alpha(U)$ is identified with
a linear subspace of $S^{\prime\, 0}_\alpha$ by the following
theorem.

\bt[density theorem]
 The space $S^0_\alpha$ is sequentially dense in
$S^0_\alpha(U)$ for each open cone $U\subset \oR^n$.
\et
  We call a closed cone $K\subset \oR^n$  a {\it carrier
  cone} of an analytic functional $v\in S^{\prime\,
  0}_\alpha$ or say that this functional   {\it is carried} by $K$ if
   $v$ admits a continuous extension to every space
   $S^0_\alpha(U)$, where  $U\supset K\setminus\{0\}$. This continuity
   property is equivalent to the belonging
    $v\in S^{\prime\, 0}_\alpha(K)$, where
   $S^0_\alpha(K)$ is by definition the union of the  spaces
   $S^0_\alpha(U)$ and is equipped with the inductive limit topology.
   The space $S^0_\alpha(\{0\})$ that corresponds to the degenerate cone
    $\{0\}$ consisting of the origin needs special consideration.
    This is the space of all entire functions satisfying the inequality
    $|f(z)|\le C\exp(b|z|)$, where $z=x+iy$ and the constants
    $C$ and $b$  depend on $f$.  From the above estimate of the test
   function behavior,  the property of  $v$ to be carried by a cone $K$
   may be thought of as a faster than  exponential decrease of this functional in the
   complement of  $K$.

\bt[quasi-localizability theorem] If a functional
 $v\in S^{\prime\, 0}_\alpha$ is carried by each of the closed cones
 $K_1$ and $K_2$, then it is carried by their intersection.
\et
 The standard compactness arguments then imply that there is
a smallest closed cone $K$ such that $v\in S^{\prime\,
0}_\alpha(K)$. This cone is called the  {\it quasi-support} of
$v$.

\bt[decomposition theorem]
 Every functional $v$ belonging to
$S^{\prime\, 0}_\alpha$ and carried by the union of two closed
cones $K_1$, $K_2$ allows a decomposition of the form $v=v_1+v_2$,
where $v_j\in S^{\prime\, 0}_\alpha(K_j)$, $j=1,2$. \et

When dealing with classes of analytic functionals possessing the
properties from Theorems 1--3, it is natural to call them
 {\it quasi-distributions}.

If a cone  $K$ is properly convex, i.e., its dual cone
$K^*=\{\eta:\,\eta x\geq0,\quad \forall x\in K\}$ has a nonempty
interior, then $e^{i\zeta x}\in S^0_\alpha(K)$ for all $\I\zeta$
in this interior and the Laplace transform
$$
\check {\bf v}(\zeta)=(2\pi)^{-n/2}(v,\,e^{i\zeta x})
 $$
is well defined for each $v\in S^{\prime\, 0}_\alpha(K)$.

\bt[Paley-Wiener-Schwartz-type theorem]
 Let a functional $v\in
S^{\prime\, 0}_\alpha$ be carried by a closed properly convex
 cone $K$. Its Laplace transform $\check {\bf
v}$ is analytic in the tubular domain $\oR^n+iV$, where $V$ is the interior of
 $K^*$, and satisfies the estimate
 \be
|\check {\bf
 v}(\zeta)|\,\leq\,C_{\varepsilon, R}(V')\, \exp\{\varepsilon\,
  |\I\zeta |^{-1/(\alpha -1)}\}\hspace{5mm} (\I\zeta\in V',\
|\zeta|\leq R)
 \label{(3)}
  \ee
 for all $\varepsilon, R>0$ and for each cone $V'$ such that
 $\bar V'\setminus\{0\}\subset V$.  If  $\I\zeta\rightarrow 0$
 inside a fixed   $V'$, then
$\check {\bf v}(\zeta)$ tends   to the Fourier transform  $\check
v$ of the functional $v$ in the topology of $S^{\prime\,
\alpha}_0$. Furthermore, if the cone $K$ is  properly convex, then
the Laplace transformation $v\to \check {\bf v}$ is a linear
topological isomorphism  of $S^{\prime\, 0}_\alpha(K)$ onto the
space  of functions that are analytic in $\oR^n+iV$ and satisfy
estimate~\eqref{(3)}.
 \et
 Theorem  1  was proved in  \cite{12},  Theorems 2 and 3 were  derived
 in \cite{10}, and Theorem~4 was established in \cite{11}, \cite{12}
 using the H\"{o}rmander $L^2$--estimates both for solutions of the
 system of nonhomogeneous Cauchy-Riemann equations and for solutions of the dual
 equation of this system. Analogous theorems hold for the
 functional classes $S^{\prime\, \beta}_\alpha$, $0<\beta<1$, and the proof
  is even simpler in this case (see \cite{1}, \cite{17}), but  the spaces
 $S^0_\alpha$ are of  particular importance because they are universal in
 the sense that no restrictions are imposed on the ultraviolet
 behavior of fields when we use these spaces.  It is worth noting that we can use
 the terms carrier cone and quai-support without specifying the index $\alpha$
 (which is quite natural because this index characterizes the behavior at
 infinity, not the local properties). In fact, the condition $v\in
 S^{\prime\,
 0}_\alpha(K)$ straightforwardly implies   that $v\in S^{\prime\, 0}_{\alpha'}(K)$  for
 all $\alpha'<\alpha$.  On the other hand, the following theorem was proved in
  \cite{13}.

\bt
 If $v\in S^{\prime\, 0}_\alpha\cap S^{\prime\,
0}_{\alpha'}(K)$ , where $\alpha'<\alpha$, then $v \in S^{\prime\,
0}_\alpha(K)$. \et

 An analogous statement for  $S^{\prime\,
\beta}_\alpha$ with $\beta\neq 0$  follows immediately from
Theorem~3 in \cite{1}, while for $\beta= 0$, we must again use the
H\"{o}rmander estimates. Another simple fact, which is a special
case of Theorem 2 in \cite{1}, is also  useful in what follows.

 \begin{lemma}[convolution lemma]For every  $v\in
S^{\prime\, 0}_\alpha(K)$ and for each test function $f\in
S^0_\alpha$, the convolution $(v*f)(x)=(v,f(x-\cdot)$ belongs to
the space $S^0_\alpha(C)$, where $C$ is any open cone such that
  $\bar C\setminus\{0\}\subset\complement K$.  Moreover, the mapping
 $S^0_\alpha\rightarrow S^0_\alpha(C):\,f\rightarrow v*f$ is continuous.
 \end{lemma}

This is almost evident because shifting a test function into the
interior of the complementary cone $\complement K$ means removing
it from the carrier cone $K$.

\section[Analytic wave front set and quasi-support]
{\large Analytic wave front set and quasi-support}

 It is well known (see, e.g., Sec.~I.2.b in \cite{22}) that the ultradistributions
 are embedded in the space of hyperfunctions with preservation of their supports.
  Therefore, we can use the general facts \cite{23} from  hyperfunction theory
  to analyze the analytic wave front set
 $WF_A(u)$ of an ultradistribution $u\in S_0^{\prime\, \alpha}
 (\oR^n)$. We recall that  $WF_A(u)$ is a closed subset of  the product
  $\oR^n\times (\oR^n\setminus \{0\})$, which is conic relative to the second
  variable. Its  projection  on the first space in the product is the
  smallest closed subset of $\oR^n$ outside of which $u$ is analytic. This
  subset is denoted by ${\rm sing\,supp}_A u$ and the cone in $\oR^n\setminus \{0\}$
   associated with a
  point $p\in {\rm sing\,supp}_A u$ is formed by those directions of a
``bad''  behavior of the Fourier transform of $u$ at infinity that
are responsible
  for the nonanalyticity at this point.

\begin{lemma}  If $u\in S_0^{\prime\, \alpha}(\oR^n)$ and if
  a closed cone $K\subset \oR^n$ is a carrier of the Fourier transform
   $\hat u$, then\footnote{The operator
    $u\to \hat u$ is the transpose of the test function transformation
     $$
     f(x)\to (2\pi)^{-n/2}\int e^{-ipx}f(x)\,{\rm}dx,
     $$
         where the  minus sign  in the exponent is opposite to the sign
for the    (inverse) operator $v\to \check v$  in Theorem 4.}
     \be
    WF_A(u)\subset \oR^n \times(K\setminus \{0\}).
     \label{(4)}
     \ee
\end{lemma}
 {\it Proof}. We cover $K$ with a finite number of closed properly convex cones
 $K_j$ and decompose
 \be
  \hat u=\sum \hat u_j,\quad \hat u_j\in S_\alpha^{\prime\, 0}(K_j),
 \label{(5)}
   \ee
  using Theorem 3. By Theorem 4,  decomposition~\eqref{(5)} induces a
  representation of $u$ in the form of a sum of boundary values of functions
  ${\bf  u}_j(\zeta)$ analytic in the tubular domains  $\oR^n+iV_j$, where
  $V_j$ is the interior of the cone $K_j^*$ that is the dual of $K_j$.
  These boundary values coincide with the boundary values in the sense
  of hyperfunctions (see, e.g.,  Theorem 11.5 in \cite{24}). By Theorem
   9.3.4 in \cite{23}, we have the inclusion
   \be
  WF_A(u)\subset \oR^n \times(\cup K_j^{**}\setminus\{0\}),
   \label{(6)}
   \ee
   where  $K_j^{**}=K_j$ because the cones $K_j$ are closed and convex.
   By making a refinement of the covering and shrinking it to  $K$, we
   obtain inclusion~\eqref{(4)}.

  \begin{remark} Lemma 2 improves Lemma 8.4.17 in
  \cite{23}, which states that for each tempered distribution $u\in S^\prime$,
  the inclusion  $WF_A(u)\subset \oR^n \times (L\setminus \{0\})$ holds,
  where $L$ is the limit cone of the set ${\rm supp}\,\hat u$ at infinity.

\end{remark}
   The cone $L$  is certainly a
  carrier of the restriction $\hat u|S^0_\alpha$.
   In fact,  this cone by definition consists of the limits of various
  sequences $t_\nu x_\nu$, where $x_\nu\in {\rm   supp}\,\hat u$ and
  $0<t_\nu\rightarrow 0$. If $L=\{0\}$, then the support of $\hat
  u$ is a bounded set  and $(\hat u,\,f_\nu)\rightarrow 0$ for any sequence
   $f_\nu$ that converges to zero in $S^0_\alpha(\{0\})$ because this convergence
   implies the uniform convergence  of all derivatives of
  $f_\nu$ on compact sets. If $L\ne\{0\}$, then  the set ${\rm supp}\,\hat u\setminus U$
   is compact for every open cone
   $U\supset L\setminus \{0\}$. The convergence of a sequence $f_\nu$ to zero in
   $S^0_\alpha(U)$ implies the convergence to zero in each of the norms
  $\sup\limits_{x\in U\cup B}|x^k\partial^{\,q}f(x)|$, where $B$ is any bounded
  neighborhood of ${\rm supp}\,\hat u\setminus U$, and we again conclude that
  $(\hat  u,\,f_\nu)\rightarrow 0$.

  \bt
   Let $u\in S_0^{\prime\, \alpha}(\oR^n)$ be a nontrivial
   ultradistribution whose support is contained in a properly convex cone
   $V$. Then only the whole  $\oR^n$ is a carrier cone of its
  Fourier transform  $\hat  u$.
  \et

{\it  Proof}. At first we assume that $0\in {\rm   supp}\,u$.
Then every vector
  belonging to the cone $-V^*\setminus\{0\}$ is an external normal to the
  support at the point 0.   By the  Kashiwara theorem (see
  Theorem 9.6.6 and Corollary 9.6.8 in \cite{23}), all  nonzero elements of the linear span
  of the set of external normals belong to $WF_A(u)_{p=0}$.
  The interior of $V^*$ is not empty because
  the cone $V$ is properly convex, and therefore
  this linear span  covers $\oR^n$. By Lemma 2, each carrier
  cone of  $\hat u$ must then coincide with $\oR^n$.

  We now suppose that $0\not\in {\rm supp}\,u$ and  that there exists an open cone
    $U$ such that $\bar U \ne \oR^n$ and $\hat u\in S^{\prime\, 0}_\alpha(U)$.
  We then consider the  series  of ``contracted''
   ultadistributions
    \be
   \sum^\infty_{\nu=1}c_\nu u_\nu,
    \label{(7)}
    \ee
    where
    $$
    (u_\nu,g(p)) \stackrel{\rm def}{=}\nu^{-n}(u,g(p/\nu)).
    $$
    Let   $\|\cdot\|'_{U,a,b}$ be the norm of the Banach space that
    is dual of $S^{0,b}_{\alpha,a}(U)$. If
    \be
    0< c_\nu < \left(\nu^2\|\hat
   u_{\nu}\|'_{U,\nu,\nu}\right)^{-1},
   \label{(8)}
    \ee
   then the series $\sum c_\nu \hat   u_\nu$ is convergent in every
   space $S^{\prime\, 0,b}_{\alpha,a}(U)$, and  series~\eqref{(7)}
   certainly converges in $S^{\prime\, \alpha}_0$. We let  $\tilde u$ denote its
   sum.  The coefficients $c_\nu$ can  be chosen so that the
   support of $\tilde u$ contains the point 0.  In fact, let $p_0$ be the
   point of ${\rm supp}\,u$ nearest to 0. We can assume without loss of
   generality that $|p_0|=1$. For each $\mu=1,2,\ldots$, there exists a function
   $g_\mu \in S^\alpha_0$ with support in the ball
   $|p-p_0/\mu|<(1/2)\left(1/\mu- 1/(\mu+1)\right)$  such that
   $(u_\mu,g_\mu)=1$. Note that ${\rm supp}\,u_\nu\cap{\rm
   supp}\,g_\mu=\emptyset$ for $\nu<\mu$. We sequentially specify the coefficients
   $c_\nu$ by imposing the conditions
   $$
    c_\nu|(u_\nu,g_\mu)|\le a_\mu/2^\nu\quad {\rm for}\quad
  \mu<\nu,
     $$
   in addition to (8). Then
    $$
    (\tilde  u,g_\mu)=a_\mu+\sum_{\nu>\mu}c_\nu(u_\nu,g_\mu)\ne 0
     $$
   for each $\mu$. Therefore $0\in{\rm supp}\,\tilde u$, and we have
   the situation examined above, which completes the proof.

   Theorem  6 can be reformulated as a uniqueness
  theorem, which is  used below.
  Namely, if we known that an ultradistribution $u\in
  S_0^{\prime\,
  \alpha}(\oR^n)$ has support in a properly convex cone and   $\hat u$
  is carried by a cone different from  $\oR^n$, then $u\equiv 0$.

\section[Asymptotic commutativity]{\large Asymptotic commutativity}

We now consider a finite family of fields $\{\phi_\iota\},\quad
\iota=1,\dots,I$, that are  operator-valued generalized functions
defined on the space $S^0_\alpha(\oR^4)$, $\alpha>1$, and
transform according to irreducible representations of the proper
Lorentz group $L_+^\uparrow$ or its covering group $SL(2,\oC)$. We
   adopt all the standard assumptions of the Wightman axiomatic approach
   \cite{2}--\cite{4}, except   local commutativity,  which cannot be formulated  in
   terms of the  test functions belonging to $S^0_\alpha$  because they are
   entire analytic  in  coordinate space. As usual, we let $D_0$ denote the minimal
   common invariant domain, which is assumed to be dense, of the field operators in
   the Hilbert space ${\mathcal H}$ of states, i.e., the vector subspace that is
   spanned by the vacuum state $\Psi_0$  and by various vectors of the form
    $$
    \phi_{\iota_1
   \ell_1}(f_1)\dots \phi_{\iota_n \ell_n}(f_n)\Psi_0 ,\qquad n=1,2,\dots,
    $$
   where  $f_k \in S^0_\alpha(\oR^4)$ and $\ell_k$ are the Lorentzian indices.

  \begin{definition}
  The field components
  $\phi_{\iota\, \ell}$ and  $\phi_{\iota^\prime \ell^\prime}$
 commute (anticommute) asymptotically for large spacelike separation of
 their arguments if the analytic functional
 \be
  \langle\Phi,\, [\phi_{\iota\,\ell}(x),\phi_{\iota^\prime
\ell^\prime}(x')]_{\stackrel{-}{(+)}}\Psi\rangle
  \label{(9)}
  \ee
is carried by the closed cone $\bar{\mathcal V}=\{(x,x') \in
\oR^8: (x-x')^2\geq 0\}$ for any  $\Phi, \Psi\in D_0$.
\end{definition}

 The matrix element~\eqref{(9)}  can be treated as a
 generalized function on $S^0_\alpha(\oR^8)$ because (see  \cite{12})
  this space coincides with the  $\tau_i$-completed tensor product
 $S^0_\alpha(\oR^4) \mathbin{\hat{\otimes}_i}S^0_\alpha(\oR^4)$ whose dual
 space,  by  definition of the inductive topology  $\tau_i$, is canonically
  isomorphic to the space of bilinear separately
  continuous forms on   $S^0_\alpha(\oR^4) \times S^0_\alpha(\oR^4)$.
  Such a coincidence also obtains  in the case  of the tensor product of several
  spaces; therefore the
   $n$-point vacuum expectation values of fields uniquely determine
  Wightman generalized functions ${\mathcal
  W}_{\iota_1 \ell_1,\dots,\iota_n \ell_n}\in S^{\prime\,
  0}_\alpha(\oR^{4n})$. We identify these objects in what follows, just as in the standard
  scheme \cite{2}--\cite{4}. Next we routinely define the expressions
   \be
   \int \phi_{\iota_1 \ell_1}(x_1)\dots
  \phi_{\iota_n \ell_n}(x_n)\, f(x_1,\dots,x_n)\,{\rm d}x_1\dots{\rm d}x_n
  \Psi_0 \qquad(n=1,2,\dots),
   \label{(10)}
   \ee
   where $f \in S^0_\alpha(\oR^{4n})$.  Namely, vector~\eqref{(10)} is the limit
   of the sequence
   $$
   \Psi_N= \sum_{\nu=1}^N\phi_{\iota_1
   \ell_1}(f_1^\nu)\dots \phi_{\iota_n \ell_n}(f_n^\nu)\Psi_0,
   $$
 where
     $$
     \sum_{\nu=1}^Nf_1^\nu(x_1)\dots f_n^\nu(x_n)=f_N \in S_\alpha^0
     (\oR^4)^{\otimes n}
     $$
     and $f_N\to f$ in $S^0_\alpha(\oR^{4n})$. The set of vectors of
     form~\eqref{(10)} spans a subspace  $D_1\supset D_0$ to which
     every operator $\phi_\iota(f)$ can be extended by continuity.
        The field operators comprise an
        irreducible system, which can be established in a manner
        analogous to that used in the standard Wightman framework \cite{2}.
          Let $M(f)=\phi_{\iota_1
        \ell_1}(f_1)\dots \phi_{\iota_n \ell_n}(f_n)$ be a monomial in the field
        components and  the operator $T$ implement the space-time
        translations. If a bounded operator $B$ acting in
        $\mathcal H$ commutes weakly with all the field operators, then
         $$
         \langle M(f)^*\Psi_0,\,B\Psi_0\rangle=
        \langle\Psi_0,\,BM(f)\Psi_0\rangle.
        $$
        Replacing  $M(f)$ here with the monomial
        $T(r) M(f)T^{-1}(r)=M(f(\cdot-r))$ and taking  the
        translation invariance of the vacuum into account, we
        obtain
         \be
           \langle
        M(f)^*\Psi_0,\,T^{-1}(r)B\Psi_0\rangle= \langle
        B^*\Psi_0,\,T(r)M(f)\Psi_0\rangle.
            \label{(11)}
           \ee
          This function of  $r$ is smooth and bounded by the constant
          $\|B^*\Psi_0\|\cdot\|M(f)\Psi_0\|$. Hence, its Fourier transform
          is a tempered distribution.
          By the spectral condition, the Fourier transform of the left-hand
          side of~\eqref{(11)} is supported in the closed  forward light cone
           $\bar \oV_+$, while  that of the right-hand side is supported
           in            $\bar \oV_-$.  Therefore, the support  is  the one-point
           set $\{0\}$, and  function~\eqref{(11)} is a bounded
           polynomial, i.e., a constant. The fact that the left-hand side
           of~\eqref{(11)} is independent of $r$  implies the invariance of the vector
           $B\Psi_0$ with respect to the space-time translations if we take into account
           that the monomial $M(f)$ is arbitrary and  $D_0$ is dense in $\mathcal
           H$. Because   the vacuum  is assumed to be the
           only invariant state, it follows that $B\Psi_0=\lambda\Psi_0$.
           Writing
            $$
            \langle
           M(f)^*\Phi,\,B\Psi_0\rangle=
   \langle\Phi,\,BM(f)\Psi_0\rangle=\lambda\langle\Phi,\,M(f)\Psi_0\rangle,
          $$
           where $\Phi\in D_0$, and again using the denseness of
            $D_0$  in $\mathcal H$, we obtain  $B=\lambda I$, as was to be
            proved.

            \medskip
              We replace the local commutativity axiom  with the
   {\it asymptotic commutativity} condition, which means that any two field
components either commute or anticommute asymptotically. This
condition is evidently weaker than local commutativity in the
sense that it is certainly  fulfilled for the restrictions of
local fields to $S^0_\alpha(\oR^4)$.  The standard considerations
of Lorentz covariance imply that the type of  commutation
relations depends only on the type of the participating fields,
not on their Lorentzian indices. Because of this we drop these
indices in what follows.

\medskip
We now consider the generalized function determined by the vacuum
expectation value
 \be
   \langle\Psi_0,\,
\phi_{\iota_1}(x_1)\dots\phi_{\iota_{k-1}}(x_{k-1})
  [\phi_{\iota_k}(x_k),\phi_{\iota_{k+1}}(x_{k+1})]_{\stackrel{-}{(+)}}
     \phi_{\iota_{k+2}}(x_{k+2})
  \dots\phi_{\iota_n}(x_n)\Psi_0\rangle,
  \label{(12)}
   \ee
  where the sign $-$ or $+$ corresponds to the type of commutation relation
  between the fields $\phi_{\iota_k}$ and  $\phi_{\iota_{k+1}}$.

   \begin{lemma}
   The  asymptotic commutativity condition implies
   that the  functional defined by~\eqref{(12)} on $S^0_\alpha(\oR^{4n})$
     extends continuously to the space
   $S^0_\alpha\left(\oR^{4(k-1)}\times U \times\oR^{4(n-k-1)}\right)$, where $U$ is any
   open cone in $\oR^{8}$ such that $\bar{\mathcal V}\setminus\{0\} \subset U$.
    Hence, this functional is carried by the closed cone
     $\bar{\mathcal V}_{n,k}=\oR^{4(k-1)}\times \bar{\mathcal V}
    \times\oR^{4(n-k-1)}$.
\end{lemma}
{\it    Proof}. First, we examine  the simplest nontrivial case
where $n=3$ and  $k=1$.   We then have a bilinear separately
continuous form on     $S^0_\alpha(\oR^8)\times
S^0_\alpha(\oR^4)$.  The
    asymptotic commutativity condition  implies that it remains
    separately continuous after giving   $S^0_\alpha(\oR^8)$  the topology
    induced by that of $S^0_\alpha (U)$.  It is not immediately clear
    that the extension by continuity to the latter space in the first
    argument, which is possible when the second argument is held fixed, yields
    a bilinear separately continuous form again. However, the space
    $S^0_\alpha(\oR^4)$,  being the inductive limit of Fr\'echet
    spaces, is barrelled, and we can therefore use standard facts on
    the extension of bilinear mappings.  By Theorem~III.5.2 in
    \cite{25}, the form under consideration is
    $\gB$-hypocontinuous, where $\gB$ is the family of all bounded subsets of
    the space $S^0_\alpha(\oR^8)$ under the topology induced by that of
    $S^0_\alpha (U)$. By Theorem 1, the space $S^0_\alpha(\oR^8)$  is sequentially
    dense in    $S^0_\alpha(U)$, and hence the family of closures
      of the specified  bounded subsets in
     $S^0_\alpha(U)$    covers
     $S^0_\alpha(U)$.  We now can apply Theorem
    III.5.4 in \cite{25}, which shows that the extension in question is indeed
    bilinear and separately continuous. It can therefore be identified with a
    continuous linear functional on  $S^0_\alpha(U\times \oR^4)=
    S^0_\alpha(U) \mathbin{\hat{\otimes}_i}S^0_\alpha(\oR^4)$, as was to be
    proved.

      For  $n=4$ and $k=1$, we consider
    expression~\eqref{(12)} as a bilinear form on
      $S^0_\alpha(\oR^{12})\times S^0_\alpha(\oR^4)$. The previous
      considerations show that it is separately continuous under the topology
      induced on $S^0_\alpha(\oR^{12})$ by that of $S^0_\alpha(U\times \oR^4)$.
 Repeating this reasoning and this time exploiting the
 denseness of  $S^0_\alpha(\oR^{12})$ in $S^0_\alpha(U\times \oR^4)$,
 we obtain a unique continuous extension to the
 space $S^0_\alpha(U\times \oR^8)$.  In the general case,
  Lemma 3 can be proved by induction on $n$.

\medskip
\noindent
     {\bf Corollary.}
     {\it If the asymptotic commutativity condition is
     satisfied for the domain $D_0$, then it  holds for the larger domain
      $D_1$ composed of various finite linear combinations of the vacuum and
      vectors of form~\eqref{(10)}.}

\medskip
     As an application of Theorem 6, we note that  under the asymptotic
     commutativity condition, the closure
     of the subspace  $L\subset \mathcal H$ consisting of vectors of  form~\eqref{(10)},
     with the fixed indices $\iota_1 \ell_1,\dots,\iota_n \ell_n$
      coincides with the closure of any subspace spanned by
      vectors of the same type but with  a different order of the field
      operators. We let $\Psi(f)$ denote  vector~\eqref{(10)},
       $\pi$  denote a permutation of the indices
      $1,\dots,n$, and  $\Psi_\pi (f)$ denote the vector that corresponds to
      the new arrangement of the operators. The orthogonality of a vector
       $\Phi$ to the subspace $L$ means that the functional
       $\langle\Phi,\,\Psi(f)\rangle$ is identically zero. But then
       the functional $\langle\Phi,\,\Psi_\pi(f)\rangle$ is carried by a cone
       different from $\oR^{4n}$. Hence, it is also equal to zero
       because, according to the      spectral condition,
      its Fourier transform is supported in a properly convex cone.

\section[Lorentz--invariant regularization]{\large Lorentz--invariant regularization}

  Let $u$ be a Lorentz-covariant ultradistribution defined on a test function
  space $S_0^\alpha(\oR^4)$,
  $\alpha>1$, and  taking values in a finite-dimensional vector space
   ${\mathcal E}$ on which the group  $SL(2,\oC)$ acts via a representation  $T$.
     We regularize the asymptotic behavior of $u$  at
   infinity by multiplying it with an invariant function of the
   form
    \be
    \omega(p/\mu)=\omega_0((p\cdot p)/\mu^2),
    \label{(13)}
    \ee
    where  $p\cdot p$  is the Lorentz square of the vector $p$ and
    $\omega_0\in  S^{\alpha^\prime}_0(\oR)$,  $1<\alpha^\prime<\alpha$,
    $\supp\omega_0\subset(-1,1)$, and $\omega_0(t)=1$ for $|t|\leq 1/2$.
    \par
    Because the mapping $p\to p\cdot p$ is analytic,
   the function $\omega$ belongs to the Gevrey class $C^L$ with
   $L=(n+1)^{\alpha^\prime}$ (see  Proposition   8.4.1 in  \cite{23}).  This means
   that $\omega$ satisfies the estimate
       \be
    \sup_{p\in \mathcal B} |\partial^k\omega(p)|\le C_{\mathcal B}h_{\mathcal
  B}^{|k|}k^{\alpha^\prime k}
     \label{(14)}
     \ee
    for every compact set ${\mathcal B}\subset \oR^4$.
   The constant $h_{\mathcal B}$ grows with increasing ${\mathcal
   B}$   and because of this,   the parameter   $\alpha^\prime$ must  be
   taken different from $\alpha$  and as close to unity as possible (see below).
Using~\eqref{(14)}, we can easily  verify that $\omega$
   is a multiplier for  $S_0^\alpha(\oR^4)$.  Evidently,
    $$
    u_{\rm reg}\stackrel{\rm def}{=}u\,\omega(p/\mu)\to u
   $$
    in the
    topology of $S_0^{\prime\, \alpha}(\oR^4,{\mathcal
   E})$ as $\mu\to \infty$.

  \bt  Let $u\in
  S_0^{\prime\, \alpha}(\oR^4,{\mathcal E})$ be a Lorentz-covariant
   ultradistribution and  $\omega$ be defined
  by~\eqref{(13)}. Then the regularized functional $u_{\rm reg}$ $($more
   precisely, the restriction $u_{\rm reg}|S^{\alpha^\prime}_0$$)$
   has a continuous extension to the space
  $S^{\alpha^\prime}_{\alpha-\alpha^\prime}$. In particular, if
  $\alpha>2$ and $\alpha^\prime<\alpha-1$, then the Fourier transform
  of $u_{\rm reg}$ is strictly localizable $($because
   $S_{\alpha^\prime}^{\alpha-\alpha^\prime}$ then contains functions of compact
   support$)$.
\et
  {\it Proof}. First of all we note that
  $S^{\alpha^\prime}_0$ is dense in $S_0^\alpha$ as well as in
  $S^{\alpha^\prime}_{\alpha-\alpha^\prime}$. Therefore, the desired extension
   is unique. We first consider  the simplest case where
   $u$ is Lorentz invariant. We can set $\mu=1$ without loss of generality.
   As shown in \cite{26},   the possibility of the extension
   depends on  the asymptotic behavior
   of the smoothed functional, i.e., the convolution
    $(u_{\rm reg}*g)(q)=(u_{\rm reg},g(q-\cdot))$, $g \in
   S^{\alpha^\prime}_0$.  For simplicity, we assume that
    $\supp g$ is contained in the ball $|p|<1$, set
     $q^2=q^3=0$ and use the light-cone variables $q^\pm=(q^0\pm q^1)/\sqrt{2}$.
    We also set $|q^-|\le  q^+,\quad q^+>1$ and let
    $\Lambda$ denote the boost  $p^+\to p^+/q^+$, $p^-\to q^+p^-$
    in the plane   $(p^0,p^1)$. Because
     $u$ and $\omega$ are Lorentz invariant, we have
     \be
     (u_{\rm  reg}*g)(q)=(u,g_q),\quad  \quad g_q(p)
  \stackrel{\rm def}{=}\omega(p)g(q-\Lambda^{-1}p).
   \label{(15)}
   \ee
  We now  estimate  the values of a functional
      $u\in S_0^{\prime\, \alpha}$ on test functions with support
  in the ball $|p|\le B$:
   \be
  |(u,\,g_q)|\le \|u\|_{\alpha,A,B}\|g_q\|_{\alpha,A,B},
   \label{(16)}
   \ee
  where
     \be
      \|g\|_{\alpha,A,B}= \sup_{|p|\le
 B} \,\sup_{k\in \oZ_+^n}\frac{|\partial^k g(p)|}{A^{|k|}k^{\alpha k}}
  \label{(17)}
    \ee
  by the definition of the topology of  $S_0^\alpha$. The points of
  ${\rm supp}\,g_q$ satisfy the inequalities
   $|p\cdot p|<1$ and $(p^2)^2+(p^3)^2<1$ by construction, and hence
   $|p^+p^-|<1$.  Furthermore, $|q^+-q^+p^+|<1$ and, as a consequence,
   $|p^-|<1/(1-1/q^+)$.  Therefore, ${\rm supp}\,g_q$ is contained in the ball of
  radius 2 if $q^+$ is large enough.  In estimating  the derivatives we take
  into account that the norm  $\|\cdot\|_{\alpha^\prime,a,1}$  with a
    suitable $a$ is finite for the function $g$ and that the transformation
    $\Lambda^{-1}$ contracts  the graph of the
    function in the variable $p^+$ in $q^+$ times. Because the change to the
   light-cone variables, as well as any linear transformation of coordinates,
   is an automorphism of the Gelfand-Shilov spaces, we obtain
   \be
   |\partial^kg(q-\Lambda^{-1}p)|\le\sup_{p\in
   \oR^4}|\partial^kg(\Lambda^{-1}p)|\le \|g\|_{\alpha^\prime,a,1}(a^\prime
    q^+)^{|k|}k^{\alpha^\prime k},
     \label{(18)}
     \ee
    where $a^\prime$ differs from $a$ by a number factor.  (It is
    easy to verify that $a^\prime\le2^{2\alpha^\prime+1}a$, but this
    refinement is unessential in what follows.) Inequalities~\eqref{(14)}
     and \eqref{(18)} combined with the Leibnitz formula give
     $$
    |\partial^kg_q(p)|\le C\|g\|_{\alpha^\prime,a,1}(a^\prime
    q^++h)^{|k|}k^{\alpha^\prime k},
    $$
    where $h$ corresponds to the compact set  $|p|\le2$.
    Therefore,
    \begin{multline}
    |(u_{\rm reg}*g)(q)|\le C
    \|u\|_{\alpha,A,2}\,\|g\|_{\alpha^\prime,a,1}\, \sup_k \frac{(a^\prime
   q^++h)^{|k|}}{A^{|k|}k^{(\alpha-\alpha^\prime)k}}\le\\
      C^\prime\|g\|_{\alpha^\prime,a,1}
  \exp\left\{\left(a^{\prime\prime}|q|/A\right)^{1/(\alpha
  -\alpha^\prime)}\right\}.
   \label{(19)}
   \end{multline}
   In other quadrants of the plane $q^2=q^3=0$, the asymptotic behavior of
   the convolution can be  similarly estimated using
   $|q^+|$ or $1/|q^-|$ as a  boost parameter.  For $q$ ranging  a
   bounded set, the use of a boost is unnecessary and the convolution is
   obviously majorized by the constant $C^{\prime\prime}
   \|g\|_{\alpha^\prime,a,1}$.  Further, any vector $q\in \oR^4$ can be
  carried to a point $\tilde q$  of this plane by a suitable space rotation
    $R$ and $(u_{\rm reg}*g)(q) = (u_{\rm
   reg}*g_R)(\tilde q)$, where $g_R(\cdot)=g(R^{-1}(\cdot))$.
   The correspondence $g\to g_R$ is a continuous mapping from
    $S^{\alpha,a}_{0,1}$ to $S^{\alpha,\tilde a}_{0,1}$ (where
    $\tilde a\le 3a$ as is easy to see).  Hence, (19) is
   satisfied in the whole space $\oR^4$, possibly with some new constants
    instead of $C^\prime$ and $a^{\prime\prime}$.  Because $A$ can be taken
    arbitrarily large,  estimate~\eqref{(19)} implies that
    $u_{\rm reg}$ increases no faster than exponentially of order
     $1/(\alpha-\alpha^\prime)$ and type zero. By Proposition
    1 in \cite{26}, it follows that $u_{\rm
    reg}$ has an extension to the space
   $S^{\alpha^\prime}_{\alpha-\alpha^\prime}$, which can
 be determined by the formula
    \be
    (\tilde u_{\rm
   reg},\,g)=\int\,(u_{\rm reg},\,\chi_0(q-\cdot)g(\cdot))\,{\rm d}q,\quad g
   \in S^{\alpha^\prime}_{\alpha-\alpha^\prime}\,,
    \label{(20)}
    \ee
    where  $\chi_0$ is an element of
    $S^{\alpha^\prime}_0(\oR^4)$ with support in the unit ball and such that
     $\int\!\!\chi_0(p)\,{\rm d}p=1$.  Inequality~\eqref{(19)} implies that the integral
     in the right-hand side of~\eqref{(20)} does exist (see  \cite{26} for more
     details).  In the general case of a Lorentz-covariant generalized
     function $u_\ell$, the only complication is due to the
     $q$ dependence of the matrix elements
     $T_{\ell\ell^\prime}(\Lambda^{-1})$ of the representation under which
     $u_\ell$ transforms, because the analogue of formula~\eqref{(15)} involves these matrix
     elements.
  However, this dependence is polinomially bounded and has no
     effect on the exponential estimates. Theorem 7 is thus proved.

   We also need the following result for the case of several variables
   $(p_1,\dots,p_n)\in \oR^{4n}$.

   \bt
    Let $u$ be a Lorentz-covariant ultradistribution
   taking values in ${\mathcal E}$ and defined on
    $S^\alpha_0(\oR^{4n})$. If ${\rm
   supp}\,u\subset \bar \oV_+\times\dots\times \bar \oV_+=\bar \oV_+^n$ and
   the regularizing  multiplier has the form
   $\omega (p)=\omega_0 (P\cdot P)$, where $P=\sum_{i=1}^n p_i$ and
   $\omega_0\in S_0^{\alpha^\prime}(\oR)$, $1<\alpha^\prime<\alpha$, then
   the conclusion of Theorem $7$ holds  for $u_{\rm
   reg}=u\,\omega(p/\mu)$
\et
  {\it Proof}. We again consider the convolution $(u_{\rm
  reg}*g)(q)$, $g\in S_0^{\alpha^\prime}$, assuming now that
   ${\rm supp}\,g$ is contained in the set
  $\{p\in\oR^{4n}:  \,|p|<1/n\}$.  Letting $Q=\sum q_i$, setting
  $Q^2=Q^3=0$ and assuming that $|Q^-|<Q^+$, $Q^+>1$, we use
   the transformation $\Lambda:\,
  p^+_i\to p^+_i/Q^+$, $p^-_i\to Q^+p^-_i$, $i=1,\dots,n$. For the points
   at which the function $g_q=\omega(p)g(q-\Lambda^{-1}p)$
   differs   from zero, the inequalities $|P\cdot P|<1$, $(P^2)^2+(P^3)^2<1$,
  and $|Q^+-Q^+P^+|<1$ hold.  Therefore, the previous reasoning shows that
   ${\rm supp}\,g_q$  is contained in the set $|P|<2$ if
    $Q^+$ is large enough.  We fix a neighborhood
    ${\mathcal U}$ of $\supp u$ by taking the union of a
    neighborhood of the origin with the product $U^n$, where $U$ is an open
    properly convex cone in $\oR^4$ containing $\bar \oV_+\setminus
    \{0\}$.  For the points of $U^n$, the inequality
    $|p|<\theta |P|$ holds with some constant $\theta>0$ because
     otherwise we could find a sequence of points
     $p_{(\nu)}\in U^n$ such that $|p_{(\nu)}|=1$ and
    $|P_{(\nu)}|<1/\nu$.  Then we could choose a convergent subsequence
    whose limit $\bar p$ is a nonzero vector in  $\bar
    U^n$ such that $|\bar P|=0$,  contradicting the assumption that the
    cone $U$ is properly convex.  Therefore, the set
    ${\rm supp}\,g_q\cap {\mathcal U}$ lies in the ball of radius
    $2\theta$.  Let $\chi$ be a multiplier for
    $S^{\alpha'}_0$ equal to unity in a neighborhood of $\supp u$
    and zero outside ${\mathcal U}$. Then
    $(u,\,g_q)=(u,\,\chi g_q)$.  It remains to estimate the norm
    $\|\chi g_q\|_{\alpha,A, 2\theta}$. This reduces to a minor
    modification of the arguments used in the derivation of
    inequality~\eqref{(19)} and yields the same result,
    which completes the proof.

  \section[The role of Jost points in  nonlocal field theory]
  {\large The role of Jost points in  nonlocal field theory}

In  local field theory \cite{2}--\cite{4}, the real points of the
extended domain of analyticity of the Wightman functions
${\mathcal W}(x_1,\dots,x_n)$ are referred to as Jost points. The
Bargman-Hall-Wightman theorem shows that this extension is
obtainable by applying various complex Lorentz transformations to
the primitive domain of analyticity determined by the spectral
condition. In terms of the difference variables
$\xi_k=x_k-x_{k+1}$, $k=1,\dots,n-1$, on which the Wightman
functions actually depend because of the translation invariance,
the set of Jost points is written as
 \be
 {\oJ}_{n-1}=\left\{\xi\in \oR^{4(n-1)}\colon \left(\sum
\lambda_k\xi_k\right)^2< 0\quad \forall\, \lambda_k \geq 0,\, \sum
\lambda_k \ne 0\right\}.
 \label{(21)}
 \ee
We let ${\mathcal J}_n$  denote the inverse image of this open
cone in $\oR^{4n}$.
     If the vacuum expectation values grow faster than
 exponentially of order one in momentum space, then the domain of
 analyticity in coordinate space is empty because such  growth
 is incompatible with the Laplace transformation.
 However,  L\"ucke  observed \cite{7} that
 the Jost points still play an important part in this essentially
 nonlocal case.  In this section, we  prove that for
 arbitrary high-energy behavior, the complement of the Jost cone is a
 carrier for some combinations of vacuum expectation values arising when
  deriving the  spin-statistics relation and
  PCT symmetry.

  We first   pass
  to the difference variables, which  requires  a slightly more
involved argument than in the standard theory \cite{2}--\cite{4}
of tempered fields.

\begin{lemma}
 For every translation-invariant functional
${\mathcal W}\in S^{\prime\, 0}_\alpha(\oR^{4n})$, there exists a
functional $W \in S^{\prime\, 0}_\alpha(\oR^{4(n-1)})$ such that
  \be
  ({\mathcal W},\,f) = \left(W, \int\!f_t(\xi)\,{\rm
d}\xi_n\right),\quad {\rm where}\,\, f_t(\xi)= f(\xi_1+\dots
+\xi_n, \xi_2+\dots +\xi_n,\dots, \xi_n).
  \label{(22)}
  \ee
  The condition  $W \in S^{\prime\,
0}_\alpha(U)$, where $U$ is an open cone in $\oR^{4(n-1)}$,
amounts to the condition ${\mathcal W} \in S^{\prime\,
0}_\alpha({\mathcal U})$, where ${\mathcal U}=\{x\in
\oR^{4n}\colon (x_1-x_2,\dots,x_{n-1}-x_n)\in U\}$.
\end{lemma}
{\it Proof.} The linear transformation \be
 t:\quad
(x_1,\dots,x_n)\rightarrow(\xi_1=x_1-x_2,\dots,\xi_{n-1}=x_{n-1}-x_n,
\xi_n=x_n),
 \label{(23)}
 \ee
that takes each test function $f(\xi)$ to $f_t(\xi)=f(t^{-1}\xi)$, is
an automorphism of $S^0_\alpha(\oR^{4n})$, and the integration over
 $\xi_n$ maps this space  continuously onto $S^0_\alpha(\oR^{4(n-1)})$.
 Hence, assigning a functional ${\mathcal W}$ to each
 $W\in S^{\prime\, 0}_\alpha(\oR^{4(n-1)})$ by  formula~\eqref{(22)},
 we obtain an injective mapping $S^{\prime\,
 0}_\alpha(\oR^{4(n-1)})\rightarrow S^{\prime\, 0}_\alpha(\oR^{4n})$, which is
evidently continuous under the weak topologies of the dual spaces.
 Lemma 4 asserts that every translation-invariant  functional
 ${\mathcal W}$ belongs to the range of this mapping. In fact, its
 regularization through the convolution by a  $\delta$-function-like sequence of test
 functions yields a sequence ${\mathcal W}_\nu$ for which
 representation~\eqref{(22)} is obviously valid with smooth functions $W_\nu$.
   The sequence $W_\nu$
 is weakly fundamental, and because $S^{\prime\, 0}_\alpha$ is a Montel
  space,\footnote{Instead of ``Montel space,'' which is currently conventional,
   the term ``perfect space'' was used in \cite{21}.}
    it converges to a functional $W$ whose image is ${\mathcal W}$.

 The second conclusion of Lemma 4 is also evident because
   transformation~\eqref{(23)} converts the subspace of functionals
 invariant with respect to simultaneous translation in all variables
  into the subspace of functionals invariant with respect to
 translation in the last variable  and because the indicator
 function in the definition of $S^{\prime\,
 0}_\alpha({\mathcal U})$ can be taken in the multiplicative form, i.e., as the
 product of a function depending on $\xi_n$ and a function of the remaining
 variables $\xi_k$ (see Sec.~3 in \cite{12}), which completes the proof.
 \par
    Let $K$ be a carrier cone of  $W$. It follows from Lemma 4
  that its inverse image $\mathcal K$ in
 $\oR^{4n}$ is a carrier of $\mathcal W$.
 In fact, if an open cone $U$ contains the cone $K\setminus
 \{0\}$ and shrinks to it, then $\mathcal U$ is contained in any given
 conic neighborhood of $\mathcal K$.

\begin{theorem}
 Let $\phi$ be a field defined on the  space
$S^0_\alpha(\oR^4)$, $\alpha>2$, and transforming according to an
irreducible representation of the group $SL(2,\oC)$.  Let
${\mathcal W}(x_1,x_2)$ denote the  generalized Wightman function
determined by the vacuum expectation value $\langle\Psi_0,\,
\phi(x_1)\phi^*(x_2)\Psi_0\rangle$. If
 $\phi$ has an integer spin, then, as a consequence of the
Poincar\'e covariance and the spectral condition,  the difference
 \be
 {\mathcal W}(x_1,x_2)-{\mathcal W}(x_2,x_1)
 \label{(24)}
\ee
 is carried by the cone $\bar {\mathcal V}=\complement {\mathcal J}_2$. In
the case of half-integer spin, this cone is a carrier of the sum
 \be
 {\mathcal W}(x_1,x_2)+{\mathcal W}(x_2,x_1).
 \label{(25)}
 \ee
\end{theorem}
{\it Proof.} Lemma 4 reduces the problem to the derivation of the
corresponding properties for the functional $W \in S^{\prime\,
0}_\alpha(\oR^4)$. We apply the ultraviolet regularization
described above and use the notation
 $\check W_{\displaystyle \mu}(p)=\check W(p)\,\omega(p/\mu)$, where
 $\omega$ is chosen as in Sec. 5.
 Because of the spectral condition, ${\rm supp}\,{\check W}_{\displaystyle
 \mu}$ lies in the cone $\bar \oV_+$ and by Theorem 7,
 the functional ${\check W}_{\displaystyle \mu}$ is defined
 on the space $S^{\alpha^\prime}_{\alpha-\alpha^\prime}$, where
 $\alpha-\alpha^\prime >1$, i.e., the growth of ${\check W}_{\displaystyle
\mu}$ at infinity is no worse than exponential of an order less
than one.  Therefore, ${\check W}_{\displaystyle \mu}$ has an
(inverse) Laplace transform ${\bf W}_{\displaystyle \mu}(\zeta)$
holomorphic in the usual tubular domain $\oT_+=\oR^4-i\oV_+$,
whose boundary value is
 $W_{\displaystyle \mu}\in
 S_{\,\,\alpha^\prime}^{\prime\,\alpha-\alpha^\prime}$ (see, e.g.,
Theorem 4 in \cite{26}, where details of the extension of the
 Paley-Wiener-Schwartz theorem to the
 generalized functions  of this class were set forth).
 Because the regularization preserves the Lorentz covariance, we can apply
  the Bargman-Hall-Wightman theorem  \cite{2}--\cite{4}
   to ${\bf W}_{\displaystyle \mu}(\zeta)$,
     which shows that this function allows an analytic continuation to
 the extended domain $\oT_+^{\,\rm ext}$ and the continued function is
 covariant under the complex Lorentz group $L_+(\oC)$.  For the field
 combination $\phi\phi^*$ in question, the transformation properties of
 the analytic Wightman function under the space-time reflection
 $PT\in L_+(\oC)$ are
  \be
  {\bf W}_{\displaystyle \mu}(\zeta)=\pm {\bf
 W}_{\displaystyle \mu}(-\zeta),
  \label{(26)}
  \ee
 where (from here on) the upper and lower signs correspond to the respective fields with
 integer and half-integer spins. This transformation law is the
 basic point, just as in the classical proof of the spin-statistics
 theorem \cite{2}--\cite{4}.  Since $\oT_+^{\,\rm ext}$ contains all spacelike points,
  relation~\eqref{(26)} implies that the generalized function $F_{\displaystyle
 \mu}\stackrel{\rm def}{=} W_{\displaystyle \mu}(\xi)\mp W_{\displaystyle
 \mu}(-\xi)$ has support in the closed light cone ${\bar \oV}$ and
 therefore  allows a continuous extension to the space $S^0_{\alpha^\prime}(\oV)$. In
 fact, this extension can be defined by the formula $(\tilde
 F_{\displaystyle \mu},f)=(F_{\displaystyle \mu},\chi f)$, where $\chi$ is a
  multiplier for $S_{\alpha^\prime}^{\alpha-\alpha^\prime}$, which is
  identically equal to unity in an $\epsilon$-neighborhood of $\bar \oV$ and
  vanishes outside the $2\epsilon$-neighborhood. Such a multiplier satisfies
  the estimate
  \be
   |\partial^q\chi(x)|\le C
 h^{|q|}q^{(\alpha-\alpha^\prime)q},
 \label{(27)}
  \ee
 while the derivatives of any function $f \in
 S^0_{\alpha^\prime}(\oV)$ satisfies the inequalities
  $$
  |\partial^qf(x)|\le
 C_\epsilon\,\|f\|_{a,b}\,b^{|q|}\exp\{-|x/a|^{1/\alpha^\prime}\}
 $$
  on its support, as is easy to verify using the Taylor formula. Hence,
   the multiplication by $\chi$ continuously maps  $S^0_{\alpha^\prime}(\oV)$ into
 $S_{\alpha^\prime}^{\alpha-\alpha^\prime}$.
   It is  important
 that the extensions  $\tilde F_{\displaystyle \mu}$ are compatible with each
 other if   $\mu$ and  $\mu^\prime$ are large enough compared to $b$,
 namely,
   \be
  {\tilde  F}_{\displaystyle
 \mu}|S_{\alpha^\prime,a}^{0,\,b}(\oV)={\tilde F}_{{\displaystyle
  \mu}^\prime}|S_{\alpha^\prime,a}^{0,\,b}(\oV).
  \label{(28)}
  \ee
 To prove this claim, we note that $(W_{\displaystyle \mu},f)=(W,f)$ for $f\in
S_{\alpha^\prime,\,a^\prime}^{0,\,{\displaystyle \mu}/4}$ at
arbitrary $a^\prime$. In fact,  for such a function, we have
$$
|f(z)|\le
   C\exp\left\{-|x/a^\prime|^{1/\alpha^\prime}+(\mu/4)\sum|y_i|\right\}
   $$
    and this
   estimate implies (again by the Paley-Wiener-Schwartz theorem, this time in
   its simplest version dealing with functions in $\mathcal D$; see, e.g.,
   Theorem 7.3.1 in \cite{23}) that  $\supp\check f$ is contained in the
   ball $|p|\le \mu/2$, where $\omega(p/\mu)=1$ by the construction
   in   Sec.~5.  The detailed formulation of Theorem 1 given in
   \cite{12}    shows that there is a constant $c$ such that for
   $b^\prime>cb$ and $a^\prime>ca$, the space
   $S^{0,b^\prime}_{\alpha^\prime,\,a^\prime}$ is dense in
   $S^{0,b}_{\alpha^\prime,\,a}(\oV)$ in the topology of
$S^{0,b^\prime}_{\alpha^\prime,\,a^\prime}(\oV)$. Hence, equality
 (28) is satisfied for $\mu, \mu^\prime>4cb$.  Therefore,  the nonregularized
functional $W(\xi)\mp W(-\xi)$  also has a continuous extension to
 $S_{\alpha^\prime}^0(\oV)$.  Applying  Theorem 5 and Lemma  4, we
  complete the proof.

\medskip
    Theorem 9 is a special case of the following more general
    statement.

\bt Let $\{\phi_\iota\}$  be a family of fields that
 are defined on the test function space
$S^0_\alpha(\oR^4)$, $\alpha>2$, and transform according to
irreducible representations $(j_\iota,k_\iota)$ of the group
$SL(2,\oC)$.  Let  ${\mathcal W}_{\iota_1\dots \iota_n}$ be the
Wightman function determined by the $n$-point vacuum expectation
value $\langle\Psi_0,\, \phi_{\iota_1}(x_1)\dots
\phi_{\iota_n}(x_n)\Psi_0\rangle$. The cone $\complement{\mathcal
J}_n$ complementary to the Jost cone
 is a carrier of the generalized function
   \be
   {\mathcal
W}_{\iota_1\dots \iota_n}(x_1,\dots,x_n)-(-1)^{2J}{\mathcal
W}_{\iota_1\dots
  \iota_n}(-x_1,\dots,-x_n),
   \label{(29)}
   \ee
  where
 $J=j_{\iota_1}+\dots+j_{\iota_n}$.
\et
 The proof is completely analogous to that given above, with the only
 difference that we now  use a cutoff function of the form
  $ \omega(P/\mu)$, where
 $P=p_1+\dots+p_{n-1}$ and $p_k$ are  conjugate with $\xi_k=x_k-x_{k+1}$,
  then appeal to Theorem 8 instead of Theorem 7, and apply
 the familiar transformation law of the  $n$-point analytic Wightman
 function of irreducible fields with respect to the space-time
 reflection.

\section[Generalization of the spin-statistics theorem]
{\large Generalization of the spin-statistics theorem}

We begin by deriving an analogue of the Dell'Antonio lemma, which
shows that each pair of nonzero local fields $\phi$, $\psi$ has
the same type of commutation relations as the pair $\phi$,
$\psi^*$.

\bt Let the fields $\phi$, $\psi$, and their Hermitian adjoints be
defined on the test function space $S^0_\alpha(\oR^4)$,
$\alpha>1$.  If $\phi$ has different asymptotic
 commutation relations with $\psi$ and $\psi^*$, then either
$\phi(x)\Psi_0=0$, or $\psi(x)\Psi_0=0$.
\et

{\it Proof.} For definiteness, we assume that $\phi$ commutes
asymptotically with $\psi$ and anticommutes asymptotically with
 $\psi^*$ for large spacelike separations of the arguments.
We  consider the following sum of vacuum expectation values:
 \begin{multline}
\langle\Psi_0,\, \phi^*(x_1) \phi(x_2)\psi^*(y_1)
\psi(y_2)\Psi_0\rangle + \langle\Psi_0,\, \phi^*(x_1) \psi^*(y_1)
\psi(y_2)\phi(x_2)\Psi_0\rangle =\\
 = \langle\Psi_0,\, \phi^*(x_1) [\phi(x_2),\psi^*(y_1)]_+
\psi(y_2)\Psi_0\rangle + \langle\Psi_0,\, \phi^*(x_1)
\psi^*(y_1)[\psi(y_2),\phi(x_2)]_-\Psi_0\rangle.
 \label{(30)}
 \end{multline}
By Lemma 3, this  functional is carried by the union of the cones
$\{(x,y)\colon(x_2-y_1)^2\geq 0\}$ and $\{(x,y)\colon
(x_2-y_2)^2\geq 0\}$. We average it with a test function of the
form
 $$
 \bar f(x_1)f(x_2)\bar g(y_1-\lambda r) g(y_2-\lambda r),
 $$
  where
$r$ is a fixed spacelike vector and $\lambda>0$.  The result of
averaging is  a convolution considered on the ray
 $x_1=x_2=0$, $y_1=y_2=\lambda r$, and by Lemma 1,
it decreases as $\lambda \rightarrow \infty$ because this ray does
not belong to the carrier cone.  On the other hand, just as in the
original reasoning of Dell'Antonio, it can be written in the form
  \be
    \langle\Psi_0,\,
\phi(f)^* \phi(f)T(\lambda r)\psi(g)^* \psi(g)\Psi_0\rangle +
\|\psi(g)T^{-1}(\lambda r)\phi(f)\Psi_0\|^2 ,
 \notag
    \ee
   where $T(\lambda r)$ implements space-time translations. As $\lambda \rightarrow
   \infty$, the first term  of the sum tends to
$$
\|\phi(f)\Psi_0\|^2\|\psi(g)\Psi_0\|^2
$$
 by the cluster
decomposition property, which can be derived from the Wightman
axioms without using locality (see \cite{2}, \cite{3}). Therefore,
if $\psi(g)\Psi_0\neq 0$ for at least one test function $g$, then
$\phi(f)\Psi_0=0$ for all $f \in S^0_\alpha(\oR^4)$, which
completes the proof.

 \bt  Let  $\phi$  be a field that is defined on
 the  space $S^0_\alpha(\oR^4)$ with the index
 $\alpha>2$ and  transforms according to an
irreducible representation  of the group $L(2,\oC)$. The anomalous
asymptotic commutation relation between $\phi$ and its adjoint
$\phi^*$ $($that is, anticommutativity in the case of integer spin
and  commutativity in the case of half-integer spin$)$ implies
that $\phi(f)\Psi_0=\phi^*(f)\Psi_0=0$ for all $f \in
S^0_\alpha(\oR^4)$.
 \et

 {\it   Proof.}  Suppose   $\phi$ is an integer spin field. The anomalous
     commutation relation would imply, in particular, that the sum
      \be
      \langle\Psi_0,\,
      \phi^*(x_1)\phi(x_2)\Psi_0\rangle+ \langle\Psi_0,\,
      \phi(x_2)\phi^*(x_1)\Psi_0\rangle
       \label{(31)}
       \ee
      is carried by the cone $\bar{\mathcal V}$. Then Theorem 9 shows that
       this cone is also a carrier of the sum
        \be
        \langle\Psi_0,\,
      \phi^*(x_1)\phi(x_2)\Psi_0\rangle+ \langle\Psi_0,\,
      \phi(x_1)\phi^*(x_2)\Psi_0\rangle.
       \label{(32)}
        \ee
        In momentum space, both of the vacuum expectation values
        in~\eqref{(32)} have support in the properly convex cone
       $\{p\in \oR^8\colon p_1+p_2=0,\quad p_1\in \bar \oV_+\}$. Therefore,
        generalized function~\eqref{(32)} is equal to zero by
        Theorem~6. Averaging it with  $\bar f(x_1)f(x_2)$,
       we get
        $$
        \|\phi(f)\Psi_0\|^2
       +\|\phi^*(f)\Psi_0\|^2=0.
        $$
       In the case of half-integer spin, the reasoning is the same
       with a proper change of the signs in the formulas. Theorem 12 is
       proved.

       \medskip
\noindent
      {\bf Corollary.} {\it  In any field theory satisfying
      the asymptotic      commutativity condition with test
      functions in $S^0_\alpha(\oR^4)$, $\alpha>2$, the  equality
       \be
      \langle\Psi_0,\, \phi^*(x_1)\phi(x_2)\Psi_0\rangle= \langle\Psi_0,\,
      \phi(x_1)\phi^*(x_2)\Psi_0\rangle.
      \label{(33)}
       \ee
holds.

\medskip
      Proof.} The difference of these vacuum expectation values can
      be written as
      \be
       \langle\Psi_0,\,[\phi^*(x_1)\phi(x_2)]_\mp\Psi_0\rangle \pm
      \langle\Psi_0,\, \phi(x_2)\phi^*(x_1)\Psi_0\rangle - \langle\Psi_0,\,
      \phi(x_1)\phi^*(x_2)\Psi_0\rangle.
       \label{(34)}
       \ee
      By Theorem 9,  expression~\eqref{(34)} is carried by the cone $\bar{\mathcal V}$
      for both  integer and half-integer spin cases, and Theorem 6
      shows that this property is compatible with the spectral condition
      only if~\eqref{(33)} is satisfied.

      \bt
       If in a field theory  satisfying the asymptotic
      commutativity condition  with test
      functions in $S^0_\alpha(\oR^4)$, $\alpha>2$, we have
      $\phi(f)\Psi_0=0$ for all test functions, then the field
       $\phi$ vanishes.
\et
   {\it Proof.} It follows from the assumptions of the theorem that all
      vacuum expectation values involving at least one operator $\phi$
      vanish.  For instance, if $\phi$ stands in the next to the last
      position, then the vacuum expectation value
    $$
      \langle \Psi_0,\,\phi_{\iota_1},\dots,\phi_{\iota_{n-1}}\phi\,
   \phi_{\iota_n}\Psi_0\rangle
$$
  coincides with the generalized function
 $$
      \langle
      \Psi_0,\,\phi_{\iota_1},\dots,\phi_{\iota_{n-1}}[\phi,
      \phi_{\iota_n}]_\mp\Psi_0\rangle
       $$
   carried by the cone $\oR^{4(n-1)}\times \bar
   {\mathcal V}$, while the support of its Fourier transform lies in the
   properly convex cone determined by the spectral condition. Hence,
   it vanishes by Theorem 6. Next, we use the induction argument.
   Taking cyclicity of the vacuum into account, we obtain
   $\langle\Phi,\phi(f)\Psi\rangle=0$ for all $\Phi\in\mathcal H$,
   $\Psi\in D_0$, and $f\in S^0_\alpha(\oR^4)$.  The operator $\phi(f)$ is
   closable because its adjoint is densely defined. Hence $\phi(f)=0$, which
   completes the proof.

\medskip
   The reasoning above remains valid if $\phi$ is replaced
   with any monomial  $M$ in the field components. Therefore,
   $M\Psi_0=0$ implies $M=0$, which equally follows from
 Theorem~13 and the cluster decomposition property.

 We also need  the following simple assertion whose proof is
 similar to the proof of Theorem 11.

\bt  Let $\{\phi_\iota\}$ be a family of fields defined on the
space $S^0_\alpha(\oR^4)$, $\alpha>1$. If two monomials
$M=\phi_{\iota_1}(x_1)\dots \phi_{\iota_m}(x_m)$ and
$N=\phi_{\iota_1^\prime}(y_1)\dots \phi_{\iota_n^\prime}(y_n)$
anticommute asymptotically for large spacelike separations of the
set of points
 $(x_1,\ldots,x_m)$ from the set of points $(y_1,\ldots, y_n)$, then either
  $\langle\Psi_0,\,M\Psi_0\rangle\equiv0$ or
$\langle\Psi_0,\,N\Psi_0\rangle\equiv0$. \et

{\it Proof.}  In this case, we have the generalized function
 $$
\langle\Psi_0,\,M(x_1,\dots,x_m)N(y_1,\dots,y_n)\Psi_0\rangle +
\langle\Psi_0,\,N(y_1,\dots,y_n)M(x_1,\dots,x_m)\Psi_0\rangle
$$
carried by the cone
$$
\bigcup_{\stackrel{k=1,\dots,m}{l=1,\dots,n}}\{(x,y)\in
   \oR^{4(m+n)}\colon (x_k-y_l)^2\geq 0\}.
  $$
  Averaging it with a test function of the form
   $f(x_1,\dots,x_m)g(y_1-\lambda r,\dots,y_n-\lambda r)$, as
  before,  we obtain a smooth function of the parameter $\lambda$, which
  decreases rapidly as $\lambda \rightarrow \infty$ by Lemma 1.
 On the other hand,  this function can be written as
  $$
  \langle\Psi_0,\,M(f)T(\lambda r)N(g)\Psi_0\rangle +
   \langle\Psi_0,\,N(g)T^{-1}(\lambda r)M(f)\Psi_0\rangle,
  $$
  where both terms in sum tend to $\langle\Psi_0,\,M(f)\Psi_0\rangle
 \langle\Psi_0,\,N(g)\Psi_0\rangle$ by the cluster property.
 Therefore, this product of vacuum expectation values is equal to
  zero for all $f \in
 S^0_\alpha(\oR^{4m})$ and $g \in S^0_\alpha(\oR^{4n})$. Theorem 14 is thus
 proved.

      We can now derive an analogue of the Araki theorem
      on the reduction of  commutation relations to the normal form.
      We follow the usual stipulation \cite{2} that the family under
      consideration does not include fields that vanish identically  and
      that for every non-Hermitian field  $\phi_\iota$, its
      adjoint   $\phi^*_\iota$ enters in this family with some index
       $\bar \iota\neq \iota$.

     \bt In any theory of Wightman fields
      $\{\phi_\iota\}$, $\iota=1,\dots,I$, satisfying the
     asymptotic commutativity condition with test functions in
       $S^0_\alpha(\oR^4)$,  $\alpha>2$, there exists a Klein transformation
     $\phi_\iota \Rightarrow \phi_\iota^\prime$, which reduces the
     commutation relations to the normal form, that is, the fields
      $\phi_\iota^\prime$ of integer spin commute asymptotically with
      any field in  the new family for large spacelike
      separations of the arguments, and the transformed
      fields of half-integer spin anticommute asymptotically with each
      other. Furthermore, the fields   $\phi_\iota^\prime$
     satisfy all the other Wightman axioms, and the Hermitian
     conjugation condition
     $\phi_\iota^{\prime*}=\phi^\prime_{\bar \iota}$ holds.
\et

{\it Proof.} Theorems 11--14 reduce the proof of this statement to
an almost literal repetition of the classical derivation
\cite{2}--\cite{4} of the Araki theorem.

Let $F_\iota$ denote the spinorial number of $\phi_\iota$,
$$
 F_\iota=\begin{cases}
 0,&\text{if $j_\iota+k_\iota$ is  integer,}\\
 1,&\text{if $j_\iota+k_\iota$ is half-integer.}
\end{cases}
$$
 The asymptotic commutativity condition states that for each
pair of fields $\phi_\iota, \phi_{\iota'}$ belonging to the family
under consideration, the matrix elements of the combination
 \be
 \phi_\iota(x)\phi_{\iota'}(y)-(-1)^{F_\iota
F_{\iota'}+\,
\omega_{\iota\,\iota'}}\phi_{\iota'}(y)\phi_\iota(x),
\label{(35)}
\ee
 are carried by the cone $\bar {\mathcal V}$. Here,
$\omega_{\iota\,\iota'}=0$ if the commutation relation is normal
and $\omega_{\iota\,\iota'}=1$ otherwise.  The matrix $\omega$
 is symmetric and possesses the properties
 \be
  \omega_{\iota\,\iota'}= \omega_{\bar \iota\,\iota'}=\omega_{\iota\,\bar
 \iota'},\quad \omega_{\iota\,\iota}=0
  \label{(36)}
  \ee
  by Theorems 11 and 12. It determines the sign in the commutation relation
  for any two monomials  in fields. This sign depends
   only on how many fields of
  each type  occur in the monomials, more precisely, on the parity
  of these numbers.
  This characteristic of a monomial $M$ can be written as a
  row $\gm=(\gm^1,\dots, \gm^I)$, where
  $\gm^\iota=0$ if  $M$ contains an even number of $\phi_\iota$  and
  $\gm^\iota=1$ if this number is odd. The set $\gM$ of such rows
  whose components are  0 or 1 has the structure of a vector space over the field
   $\oZ_2$ and the mapping $M\rightarrow \gm$
  is consistent with this structure in the sense that
   \be
  \gm(M_1M_2)=\gm(M_1)+\gm(M_2).
 \label{(37)}
\ee
  The system of spinor numbers generates the linear form
   $F(\gm)=\sum F_\iota\gm^\iota$ and the matrix  $\omega$ generates
   the bilinear form $(\gm_1,\gm_2)=\sum
  \omega_{\iota\,\iota'}\gm_1^\iota\gm_2^{\iota'}$ on $\gM$.  In this notation,
the sign factor in the commutation relation for monomials $M_1$
and $M_2$ becomes
 $$
 (-1)^{F(\gm_1)F(\gm_2)+(\gm_1,\,\gm_2)},
 $$
and  properties (36) can be written as
 \be
 (\gote_\iota,\gm)=(\gote_{\bar \iota},\gm),\quad (\gm,\gm)=0,
\label{(38)}
 \ee
where $\gote_\iota=\gm(\phi_\iota)$. The second formula
from~\eqref{(38)} implies that the form $(\cdot, \cdot)$ induces a
symplectic structure on $\gM$. In particular, it implies the
identity $(\gm_1,\gm_2)+(\gm_2,\gm_1)=0$, which is equivalent to
 $(\gm_1,\gm_2)=(\gm_2,\gm_1)$ in the case of the field
$\oZ_2$. We let $\gA$ denote the set of those vectors in $\gM$
that correspond to the monomials whose vacuum expectation values
are not identically zero. From~\eqref{(37)} and the cluster
property, it follows that $\gA$ is a linear subspace of $\gM$. The
restriction of the form $(\cdot,\cdot)$ to $\gA$ is zero because
such monomials contain an even number of half-integer spin fields
and the relation $(\gm_1,\gm_2)=1$ would mean that $M_1$ and $M_2$
anticommute for large spacelike separation of the arguments, which
contradicts Theorem 14. Every basis
 $(\ga_1,\dots,\ga_q)$ in $\gA$ can be completed to a symplectic basis
 $(\ga_1,\dots,\ga_r; \gb_1,\dots,\gb_r;\gc_1,\dots,\gc_s)$,
  $r\geq q$, $2r+s=I$, in the whole  $\gM$.
In this basis,
  $$
(\ga_j,\gb_j)=1,\quad j=1,\dots,r,
  $$
 while all other pairings vanish. In particular,
 \be
  (\ga_j,\gm)=0  \quad {\rm for\,\, all}\quad j=1,\dots,r, \quad
\gm\in\gA.
   \label{(39)}
  \ee
 To each $\ga_j$, we can assign an operator $\theta_j$ acting in the Hilbert
 space ${\mathcal H}$. Namely, we set
   $$
  \theta_j\Psi_0=\Psi_0,\quad
 \theta_j M(f)\Psi_0= (-1)^{(\ga_j,\gm)}M(f)\Psi_0.
  $$
 These operators are
 well defined. In fact, if $M_1(f_1)\Psi_0=M_2(f_2)\Psi_0\neq~0$, then $\langle
  \Psi_0,\,M_1^*M_2\Psi_0\rangle\not\equiv 0$, i.e., $\gm(M_1^*M_2)\in \gA$
   and  relations~\eqref{(37)}--\eqref{(39)} imply $(\ga_j,\gm_1)+(\ga_j,\gm_2)=0$,
    which amounts to $(\ga_j,\gm_1)=(\ga_j,\gm_2)$.
     Further, the definition is consistent with the linear operations
     in ${\mathcal H}$ because
 the relation $M(f)\Psi_0 =M_1(f_1)\Psi_0+M_2(f_2)\Psi_0$, where all vectors
 are assumed nonzero, implies $(\ga_j,\gm)=(\ga_j,\gm_1)=(\ga_j,\gm_2)$.
 For example, if  $(\ga_j,\gm_1)$  differs from the other two scalar
 products, then again  using \eqref{(37)}--\eqref{(39)}, we obtain a contradiction
 because at least one of the vacuum expectation values
  $\langle \Psi_0,\,M_1^*M\Psi_0\rangle$ and
 $\langle \Psi_0,\,M_1^*M_2\Psi_0\rangle$ does not vanish.
 Hence, the  $\theta_j$'s can be extended to  $D_0$ by linearity.
 Relations~\eqref{(37)}--\eqref{(39)} imply that $\langle
\theta_j\Phi,\,\theta_j\Psi\rangle=\langle \Phi,\,\Psi\rangle$ for
all $\Phi,\Psi\in D_0$.  Therefore, every operator $\theta_j$ can
be uniquely extended to a unitary involution defined on the whole
of ${\mathcal H}$.   We set
   \be
  U_\iota=\prod_{j=1}^r
\theta_j^{\,(\gote_\iota,\gb_j)},\quad \iota=1,\dots,I.
  \label{(40)}
 \ee
The operators $U_\iota$ commute with each other, and their
 commutation relations with the fields are
 \be
 U_\iota\,\phi_{\iota'}=(-1)^{\sigma_{\iota\,\iota'}}\,\phi_{\iota'}\,U_\iota\,,\quad
{\rm where}\quad
\sigma_{\iota\,\iota'}=\sum_{j=1}^r(\gote_\iota,\gb_j)(\ga_j,\gote_{\iota'}).
\label{(41)}
 \ee
 The Klein transformation is defined by the formula
  \be
 \phi_\iota\Rightarrow\phi_\iota^\prime=
i^{\sigma_{\iota\,\iota}}\,U_\iota\,\phi_\iota.
 \label{(42)}
\ee
 Using~\eqref{(41)} and the equality $\omega_{\iota\,\iota'}=
\sigma_{\iota\,\iota'}+\sigma_{\iota'\iota}$,
 which holds by the definition of the symplectic basis, we deduce
 that for any $\Phi,\Psi \in D_0$,
 \begin{multline}
\langle\Phi,\,[\phi^\prime_\iota(x)\,\phi_{\iota'}^\prime(y)-
(-1)^{F_\iota
F_{\iota'}}\phi_{\iota'}^\prime(y)\,\phi_\iota^\prime(x)]\Psi\rangle=\\
 =\langle\Phi^\prime,\,[\phi_\iota(x)\,\phi_{\iota'}(y)-
(-1)^{F_\iota
F_{\iota'}+\omega_{\iota\,\iota'}}\phi_{\iota'}(y)\,\phi_\iota(x)]\Psi\rangle,
\notag
\end{multline}
 where
$\Phi^\prime=(-1)^{\sigma_{\iota'\iota}}(-i)^{\sigma_{\iota\,\iota}+
\sigma_{\iota'\iota'}}\,U_\iota\, U_{\iota'}\,\Phi \in D_0$.
Therefore, the transformed fields satisfy the normal asymptotic
commutation relations. It follows from~\eqref{(38)} and
\eqref{(41)} that the factor $i^{\sigma_{\iota\,\iota}}$ in
formula~\eqref{(42)} ensures the equality
$\phi_\iota^{\prime*}=\phi^\prime_{\bar \iota}$.  Verifying the
other Wightman axioms for the fields $\phi_\iota^\prime$ is
simple, which completes the proof.

\section[Generalization of the  PCT theorem]
{\large Generalization of the  PCT theorem}

     \bt In the field theory satisfying the asymptotic  commutativity
     condition with test functions in   $S^0_\alpha(\oR^4)$, $\alpha>2$,
  and with the normal spin-statistics relation, there exists an
      antiunitary $PCT$-symmetry operator $\Theta$. This operator leaves
      the vacuum state invariant, and if   $\phi_\iota$ transforms
      according to the $(j_\iota,k_\iota)$   representation of
      $SL(2,\oC)$, then its  transformation law under $\Theta$ is
      \be
        \Theta\,\phi_\iota(x)\,\Theta^{-1}=
       (-1)^{2j_\iota}\,i^{F_\iota}\,\phi^*_\iota(-x),
       \label{(43)}
       \ee
       where $F_\iota$ is the spinorial number of the field.
\et
   {\it   Proof}. It is well known that in terms of the Wightman functions, a
      necessary and sufficient condition for the operator $\Theta$ to exist
       is
      \be
      {\mathcal W}_{\iota_1\dots \iota_n}(x_1,\dots,x_n)=(-1)^{2J}\,i^F\,{\mathcal
     W}_{\iota_n\dots \iota_1}(-x_n,\dots,-x_1),
     \label{(44)}
      \ee
      where $J=j_{\iota_1}+\dots+j_{\iota_n}$ and
      $F=F_{\iota_1}+\dots+F_{\iota_n}$ is the number of
      half-integer spin fields in the set
    $\phi_{\iota_1},\dots,\phi_{\iota_n}$. We now write the difference of the
    right- and left-hand sides of (44) in the form
       \begin{multline}
    \left[{\mathcal W}_{\iota_1\dots \iota_n}(x_1,\dots,x_n)-(-1)^{2J}\,{\mathcal
      W}_{\iota_1\dots \iota_n}(-x_1,\dots,-x_n)\right] +\\
      {}+(-1)^{2J}\,\left[{\mathcal W}_{\iota_1\dots \iota_n}(-x_1,\dots,-x_n)-
      i^F\,{\mathcal W}_{\iota_n\dots \iota_1}(-x_n,\dots,-x_1)\right].
   \label{(45)}
       \end{multline}
      The asymptotic  commutativity condition implies that the expression
      in the first square brackets is carried by the cone
      $\bigcup_{k<l}\{x\in \oR^{4n}\colon (x_k-x_l)^2\geq0\}$, which is
  contained in the complement of the Jost cone because the Jost points are
      totally spacelike. By Theorem~10, the cone  $\complement{\mathcal J}_n$
  is a carrier of the functional in the second square brackets as well.
  In momentum space, both the generalized functions involved in (44)
  have support in the properly convex cone
  $$
  \{p\in \oR^{4n}\colon \sum_{k=1}^n
  p_k=0,\quad\sum_{k=1}^lp_k\in \bar \oV_+,\quad  l=1,\dots,n-1\}.
  $$
   Therefore,  equality~\eqref{(44)} holds identically by Theorem 6. We can now
   construct the operator $\Theta$  in the standard way. First,
    we define it on those vectors that are obtained by applying
    monomials in fields to the vacuum:
     $$
   \Theta\Psi_0=\Psi_0,\quad
  \Theta\phi_{\iota_1}(f_1)\dots\phi_{\iota_n}(f_n)\Psi_0
  =(-1)^{2J}\,i^F\,
  \phi^*_{\iota_1}(\tilde f_1)\dots\phi^*_{\iota_n}(\tilde f_n)\Psi_0,
  $$
  where $\tilde f(x)=\bar f(-x)$. It is easy to verify that
  $\Theta$ is well defined. In fact, taking into account
    that  $\phi^*_\iota$ transforms according to the
    conjugate representation $(k_\iota,j_\iota)$, we see that~\eqref{(44)} implies
  the relation
  $\langle\Theta\Phi,\,\Theta\Psi\rangle=\overline{\langle\Phi,\,\Psi\rangle}$
   for vectors of this special form. Therefore, if a vector $\Psi$
  is generated by different monomials $M_1(f_1)$ and $M_2(f_2)$, then the scalar
  product $\langle\Theta M_1\Psi_0,\,\Theta  M_2\Psi_0\rangle$ is equal
  to the squared length of either of the two vectors
  $\Theta M_1\Psi_0$ and $\Theta M_2\Psi_0$, i.e., these vectors coincide.
  Analogously, if $\Psi=\Psi_1+\Psi_2$, where all vectors are obtained by
  applying monomials to $\Psi_0$, then  $\Theta\Psi=\Theta\Psi_1+\Theta\Psi_2$.
  Therefore, $\Theta$ can be extended to $D_0$ by antilinearity.   A further
  extension by continuity yields an antiunitary operator defined on
  the whole of ${\mathcal H}$.

  A stronger formulation of Theorem 16  uses an analogue of the
  Jost-Dyson weak local commutativity condition.

   \begin{definition}  We say that the fields $\{\phi_\iota\}$
   defined on  $S^0_\alpha(\oR^4)$ satisfy
   the weak asymptotic  commutativity condition if for each
   system of indices  $\iota_1,\dots,\iota_n$,  the functional
       \be
   \langle\Psi_0,\, \phi_{\iota_1}(x_1)\dots\phi_{\iota_n}(x_n)\Psi_0\rangle
      -i^F\langle\Psi_0,\,
  \phi_{\iota_n}(x_n)\dots\phi_{\iota_1}(x_1)\Psi_0\rangle
   \label{(46)}
   \ee
   is carried by  the cone $\complement{\mathcal J}_n$ complementary to the Jost cone.
   \end{definition}

      The above consideration shows that this condition is equivalent to
      relation~\eqref{(44)}, i.e.,  the following statement is valid.

       \bt
        A field theory satisfying all  Wightman
    axioms with test functions in $S^0_\alpha(\oR^4)$, $\alpha>2$,
    but  with the possible exception of local or asymptotic
    commutativity,  has $PCT$ symmetry if and only if
    the weak asymptotic  commutativity condition is satisfied.
    \et

      Moreover, by Theorem 6, the field theory has
      PCT symmetry even if  difference~\eqref{(46)} is carried by
      the complement of a cone generated by an arbitrarily small
      real neighborhood of a Jost point. Our consideration also shows  that an
      analogue of the theorem on global nature of local commutativity is
      valid for the weak local commutativity. The most refined version
      of  this theorem is due to  Borchers and Pohlmeyer \cite{27}, who
      considered the theory of a scalar tempered field and established that a bound of
      the form
      \be
     |\langle\Psi_0,\,[\phi(x_1),\phi(x_2)]\phi(x_3)\dots\phi(x_n)\Psi_0\rangle|
     \le C_n\exp\{-\gamma|(x_1- x_2)^2|^{p/2}\}\quad   (p>1)
     \label{(47)}
       \ee
     on the behavior of the commutators at those points
      $(x_1,x_2,x_3,\dots,x_n)$ that belong to the cone ${\mathcal J}_n$
     together with $(x_2,x_1,x_3,\dots,x_n)$ results in the strict local
     commutativity.  Analogously, if all fields $\phi_\iota$ are defined on
     Schwartz's space $S$ and  functional (46) decreases in the cone
      ${\mathcal J}_n$ by an exponential law of type~\eqref{(47)}, then it actually
      vanishes everywhere in this cone.  In fact, using a partition of unity
     and taking the remark in Sec.~3 into account, we see that  the
     restriction of  functional~\eqref{(46)} to each space $S^0_\alpha(\oR^{4n})$
     is carried by the cone $\complement{\mathcal J}_n$ in this case.
      Therefore, the theory has   PCT as a symmetry and must
       satisfy the weak local commutativity condition by the usual
       PCT theorem \cite{2}--\cite{4}.

\section[Concluding remarks]{\large Concluding remarks}

The main result of this work is a rigorous proof  that  PCT
symmetry and the standard spin--statistics relation  are preserved
under replacing the microcausality axiom with the condition of a
fast decrease of the (anti)commutators at large spacelike
separations of the arguments, correctly formulated in terms of the
theory of analytic functionals. The absence of any experimental
indications of violation of these fundamental properties of
quantum physics is customary believed to be  evidence  for the
locality of interaction. On the contrary, we see that these
properties have deeper roots in the mathematical structure of
quantum field theory and are  essentially  asymptotic in
character. The established generalization of the spin-statistics
and  PCT theorems is of greatest possible accuracy because, as
already noted  by Pauli \cite{28}, allowing an exponential
decrease of the field (anti)commutators with order 1 and a finite
type implies the possibility of quantizing  the scalar field with
an abnormal relation $[\phi(x),\,\phi(y)]_+= \Delta^{(1)}(x-y)$,
where $\Delta^{(1)}$ is the even solution to the Klein-Gordon
equation, which behaves like $\exp(-m|{\bf x}-{\bf y}|)$ at
spatial infinity.

    In relation to the above remark about the theorem on the global
    nature of local commutativity, it should be emphasized that the
    asymptotic  commutativity condition,  being applied to the tempered
    fields (i.e., to their restrictions to $S^0_\alpha(\oR^4)$), does not
    amount to  naive bound~\eqref{(47)} and does not imply  local
    commutativity. Strictly speaking, this condition means a fast decrease
    not of the field commutator itself but of the result of smoothing it by
    convolution with appropriate test functions
     (see \cite{1} for more details), which seems reasonable from the
    physical standpoint.

     The theorems on carrier cones of analytic functionals
     presented in  Sec.~2 were established in \cite{10}--\cite{13}
     for applications to the covariant quantization of gauge models in
     which     singularities are of infrared  origin and for which the spaces
     $S^\beta_\alpha$ with indices $\beta<1$ are natural functional
     domains of definition of fields in the momentum
     representation  when the models are treated in a generic
     covariant gauge (see   \cite{29}). In particular, this formalism gives  a
     simple, general method of constructing Wick--ordered entire
     functions of the indefinite metric free fields in the Hilbert-Fock-Krein
     spaces \cite{30}.  In the present paper, the efficiency of the developed
     technique has been demonstrated by an example of solving classical
     problems in nonlocal quantum field theory. Actually, this formalism
      accomplishes the extension of the Wightman axiomatic
     approach to nonlocal quantum fields with arbitrary high-energy behavior.

\medskip
{\bf Acknowledgments.}
  The author is grateful to Professor V.~Ya.~Fainberg for helpful discussions.
  This work was supported in part by the Russian Foundation for Basic
  Research (Grant No. 96-01-00105) and in part by INTAS (Grant No. 96-0308).

 \end{document}